\documentstyle[epsfig,psfig]{texas}

\begin{document}

\title{Investigation on the Bimodal Distribution of the Duration of Gamma-ray Bursts 
from BATSE Light Curves}
\author{Wenfei Yu, Tipei Li, and Mei Wu}
\address{Laboratory for Cosmic Ray and High Energy Astrophysics\\
Institute of High Energy Physics\\
Chinese Academy of Sciences\\
P.O.Box: 918-3\\
Beijing, 100039\\
P. R. China\\
}
{\rm Email: yuwf@astrosv1.ihep.ac.cn}

\begin{abstract}
We have investigated the bimodal distribution of the duration 
of BATSE gamma-ray bursts (GRBs) by analyzing light curves of 
64 ms time resolution. 
We define the average pulse width of GRBs from the auto-correlation 
function of GRB profiles. The distribution of the average 
pulse width of GRBs is bimodal, suggesting that GRBs are composed 
of long-pulse GRBs and short-pulse GRBs. The average pulse width 
of long-pulse GRBs appears correlated with the peak flux, consistent 
with the time dilation effect anticipated from the cosmological 
origin of GRBs. However, the correlation between the average pulse width 
and the peak flux for the short-pulse GRBs doesn't show such 
a tendency, which needs further study with higher time 
resolution data.
\end{abstract}

\section{Introduction}
The distribution of the duration of gamma-ray bursts shows 
an indication of two distinct groups from earlier 
experiments\cite{ref1}\cite{ref2}\cite{ref3}\cite{ref4}. 
Data from Burst and Transient Source Experiment (BATSE) 
have confirmed the bimodal distribution 
of the duration of gamma-ray bursts. In terms of the parameter T90, which 
is the time interval during which the integrated counts of 
a burst go from 5\% to 95\% of the total integrated counts, 
the bursts are separated into two groups around T90 $\sim$ 
2 s \cite{ref5}. Time dilation, an evidence for 
the cosmological origin of GRBs, was found in the long GRBs \cite{ref6}. 
It is not yet known whether the two kinds of bursts are different 
or not. 

A recent study on the pulses in GRBs suggests that the duration of 
the equivalent width of each pulse and the mean duration of individual 
pulse are bimodal \cite{ref7}. In this paper, we present a different 
approach to investigate the average pulse width of GRBs.  

\section{Data Preparation and Pulse Width Definition}
Light curves of the BATSE GRBs in 4B catalogue are studied. The light 
curves are from concatenated DISCLA, PREB and DISCSC data, and were 
obtained from Compton Observatory Science Support Center (COSSC). They have been arranged into 
64 ms time bins. First we subtract BATSE background from GRB light curves. 
The BATSE background were estimated by a 5-degree polynomial. 
The total number of GRBs with visually acceptable background estimate is 
1186. Then 
we calculate the average pulse width $T_{P}$ of each GRB as follows.

\begin{figure}
\centering
\psfig{file=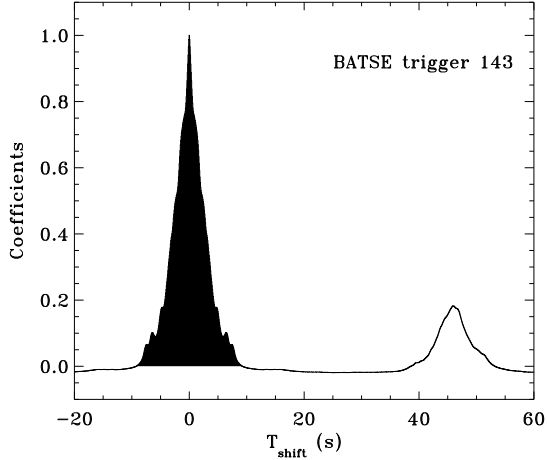,width=8cm}
\caption{The auto-correlation coefficients of BATSE trigger 143. 
The shaded area corresponds to the data used to calculate 
$T_{P}$.}\label{fig1}
\end{figure}

First we calculate the auto-correlation of the light curve of each GRB. 
The auto-correlation coefficients of the GRB, $A(i)$, are defined as follows:
$$A(i)=\frac{\sum_{k=0}^{N-i-1}(X_{k}-\bar{X})(X_{k+i}-\bar{X})}
{\sum_{k=0}^{N-1}(X_{k}-\bar{X})^2}$$
where $A(i)$ ($i=0,...,N-1$) the auto-correlation coefficient at 
$i\delta{t}$. We define the average pulse width $T_{P}$ as 
 $$T_{P}=2.0\times\sqrt{\frac{\sum_{k=1}^{M}k^2A(k)+(0.25)^2A(0)}
{\sum_{k=0}^{M}A(k)}}\delta{t}$$
where 0.25 represents the average time shift of the central bin of the 
auto-correlation coefficient A(0), and 
$M$ the maximum of $i$ with $A(i-1)$+$A(i)$ no less than 0.0 in the main 
peak of the auto-correlation. \\
The auto-correlation coefficients of BATSE trigger No.143 is shown in 
Fig.1. The data in the shaded region is used to calculate $T_{P}$. We 
calculate $T_{P}$ of each GRB and study the distribution of the average 
pulse width of the 1186 GRBs. \\

\begin{figure}
\centering
\psfig{file=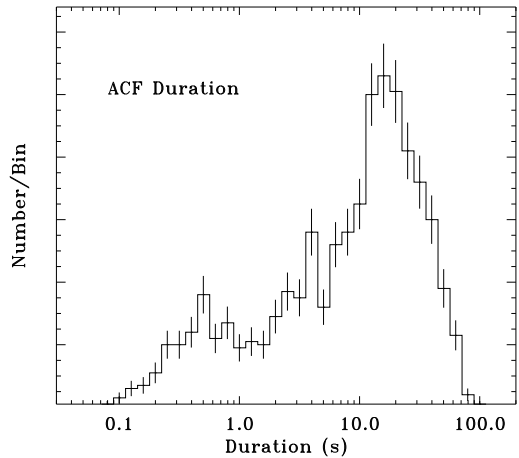,width=8cm}
\caption{Distribution of $T_{P}$. Two groups of GRBs are centered around 
0.5 s and 14 s, respectively. There is a feature around 
4 s.}\label{fig2}
\end{figure}

\section{Results}
We have obtained the following results from the study of the average 
pulse width $T_{P}$
\begin{itemize}
\item The distribution of $T_{P}$ of GRB is bimodal. This suggests 
that the average pulse width is bimodally distributed, and GRBs can be 
divided into two groups, namely shot-pulse bursts and long-pulse bursts. 
The distribution of $T_{P}$ is peaked at about 0.5 s and 14 s for the two 
groups, respectively. They are roughly separated around 2 s. This is shown in Fig.2.
\item The average pulse width of the dim long-pulse bursts are longer 
than the bright long-pulse bursts. However, the average pulse width 
of the short-pulse bursts does not show a simple relation with GRB 
peak flux. This is shown in Fig.3.
\end{itemize}

\begin{figure}
\centering
\psfig{file=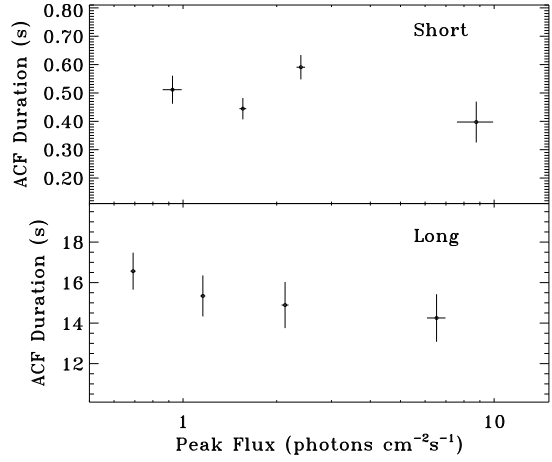,width=8cm}
\caption{$T_{P}$ vs. Peak Flux of long-pulse bursts ($T_{P}$ $>$ 2 s) 
(bottom) and short-pulse bursts ($T_{P}$ $<$ 2 s) (top). Both short-pulse 
GRBs and long-pulse GRBs are divided into 4 groups according 
to their peak fluxes. The median, instead of the mean, of $T_{P}$ 
of each group, is shown in the figure. The error on the median 
is derived from bootstrap method.}\label{fig3}
\end{figure}

\section{Summary}
We have presented our preliminary analyses of 1186 BATSE GRB 
light curves in order to study the bimodal distribution of the 
duration of GRBs. 
We conclude
\begin{itemize}
\item The duration of the average pulse width in GRBs are bimodally 
distributed. This is consistent with a different approach 
(Mitrofanov et al. 1998). 
\item Long-pulse bursts show the evidence for the time dilation effect. 
This isn't shown for the short-pulse bursts. Further study of the 
short-pulse bursts is need, and probably need to 
include correction of the BATSE selection effect and to study 
short GRBs with high time resolution TTE data.  
\end{itemize}

\section*{Acknowledgments}
WY appreciate various assistances by Dr. R. S. Mallozzi at MSFC/UAH.

\section*{References}

\end{document}